# SIMULATION OF NANOPOWDER HIGH-SPEED COMPACTION
# BY 2D GRANULAR DYNAMICS METHOD

*G.Sh. Boltachev, N.B. Volkov, A.V. Spirin, E.A. Chingina*

Institute of Electrophysics, Ural Branch of RAS, Amundsen str. 106, 620016, Ekaterinburg, Russia

**Abstract.** The paper concerns the nanopowder high-speed, $10^4 - 10^9 \ \text{s}^{-1}$, compaction processes modeling by a two-dimensional granular dynamics method. Nanoparticles interaction, in addition to known contact laws, included dispersive attraction, formation of a strong interparticle bonding (powder agglomeration) as well as the forces caused by viscous stresses in the contact region. For different densification rates, the "pressure vs. density" curves (densification curves) were calculated. Relaxation of the stresses after the compression stage was analyzed as well. The densification curves analysis allowed us to suggest the dependence of compaction pressure as a function of strain rate. It was found that in contrast to the plastic flow of metals, where the yield strength is proportional to the logarithm of the strain rate, the power-law dependence of applied pressure on the strain rate as $p \propto v^{1/4}$ was established for the modeled nanosized powders.

# 1. INTRODUCTION

The nanopowder cold compaction is a very important stage of novel nanostructured materials production by the powder metallurgy [1, 2]. As known, nanopowders in contrast to coarse-grained materials are very hard to densify due to the strong interparticle "friction", which is caused by the intense dispersion attraction, and agglomeration of particles [2, 3]. To achieve a proper compact density for sintering the high quality, defect-free ceramic article, applying the high pressure of about several gigapascals is required. Such high pressures can even exceed the durability of pressing tools [2 – 4]. Thus, the theoretical description of powder body and reliable forecasting the compaction processes take on high topicality.

The present paper is devoted to development of theoretical description of oxide nanosized powders cold compaction processes in the frameworks of the granular dynamics method [5, 6]. This method is of interest due to the oxide nanoparticles, for example, produced by the method of wires electric explosion [7] or target laser evaporation [8], usually have high strength properties and a spherical form. Therefore, such powders are the most convenient object for simulations. Nowadays the granular dynamics method is extensively used for description of compaction processes of different micro- and nanopowders [5, 6, 9, 10]. However at that quasistatic compaction processes are investigated. After every step of model cell deformation the new equilibrium locations of particles are determined during a large number of equilibration steps [5, 6].

In view of necessity to achieve extremely large compaction pressures the magnetic pulsed methods [2, 4] attract a great attention at present time. These methods allow increasing the pressure into compacts owing to the inertial effects. The relative rate of compact densification is of $10^4$ – $10^5 \, \text{s}^{-1}$. It is known that dynamical yield strength is not equal to static one, as a rule. For example, the yield strength of metal at high-speed loading can exceed the static limit by several times [11 – 13]. Corresponding studies for nanopowders have not been conducted yet.

# 2. NUMERICAL EXPERIMENT PROCEDURE

We simulate the dynamical processes of uniform pressing, which are characterized with the relative densification rate $v = (1/\rho)(d\rho/dt)$, where $\rho$ is density and $t$ is time, from value $6.8 \times 10^8 \, \text{s}^{-1}$ up to $6.8 \times 10^4 \, \text{s}^{-1}$. To perform the qualitative analysis we restrict to 2D geometry. The model cell has a form of square $L_{\text{cell}} \times L_{\text{cell}}$. The density is implied as a relative area of the model cell occupied by the particles, i.e. $\rho = (\pi/4)N_p d_g^2 / L_{\text{cell}}^2$, where $N_p = 1000$ is the number of particles in the cell, $d_g$ is the particle diameter. Periodic boundary conditions are used on all the sides of the cell. For initial packing generation, the algorithm defined in [5] is used, which allows us to create isotropic and uniform structures in a form of the connected 2D-periodic cluster. The initial density $\rho_0$ is 0.5. The system deformation is performed by simultaneous changes of cell sizes and proportional rescaling of particles coordinates. This procedure corresponds to the instantaneous propagation of elastic perturbation along the model cell. The relative displacements and rotations of particles are determined by the usual equations

$$m\frac{d^2 r}{dt^2} = f, \qquad J\frac{d^2\theta}{dt^2} = M, \tag{1}$$

where $m = (\pi/6)\rho_m d_g^3$ is the particle mass, $\rho_m$ is the density of the particle material, $f$ and $M$ are the total force and torque caused by other particles, $J = md_g^2/10$ is inertia moment, $\theta$ is the rotation angle. The Verlet algorithm [14] is applied for the numerical solve of the equation (1).

The stress tensor $\sigma_{ij}$ averaged over the model cell is calculated by the known expression [5, 9, 10]

$$\sigma_{ij} = \frac{-1}{d_g L_{\text{cell}}^2}\sum_{k<l} f_i^{(kl)} r_j^{(kl)}, \tag{2}$$

where the summation is performed over all pairs of interacting particles $(k,l)$; $f^{(k,l)}$ is the total force affecting the particle $k$ from the particle $l$; $r^{(k,l)}$ is the vector connecting the centers of the particles. The particle interactions described in detail elsewhere [5, 6] include the elastic repulsion (modified Hertz law), the "friction" forces (Cattaneo – Mindlin law), the dispersive attraction force (Hamaker's formula), and the contact elasticity of flexure because of strong interparticle bonding. Alumina is implied as the particle material for which, in particular, the Young modulus $E$ is 382 GPa and the Poisson ratio $\nu$ is 0.25. Other parameters of interaction laws correspond to the system of II type in [5, 6], which imitates strongly agglomerating nanopowders [15] with particle diameter $d_g = 10$ nm. The exception as compared to the 3D simulations is only the larger value of friction coefficient $\mu = 0.5$ used in the present study.

The high value of speed of modeled processes requires taking into account the viscous stresses in the vicinity of the contact area of particles. Using the similarity of Hooke's elastic law and the

Navier – Stokes equations the authors of [16] obtained the rigorous solution of the problem on contact interaction of viscoelastic spheres. In general case the influence of the viscous stresses has a form [16, 17]

$$f_{\text{visc}} = A \frac{df_e}{d\zeta} \frac{d\zeta}{dt},$$ \hfill (3)

where $f_{\text{visc}}$ is the total force of the viscous stresses, $f_e$ is the elastic force, $\zeta$ is the variable, which describes the body deformation, and the coefficient $A$ neglecting the bulk viscosity is described as:

$$A = \frac{\eta(1-\nu^2)(1-2\nu)}{3E\nu^2} \ .$$

The shear viscosity coefficient $\eta$ is estimated by the known data on ultrasound damping into alumina [18]. The coefficient of damping into the isotropic medium $\gamma_t = \eta\omega^2/(2\rho_m c_t^3)$ [19], where $\omega$ and $c_t$ are the frequency and speed of sound. Using the value $\gamma_t \cong 230$ dB/m at the frequency of $\omega/2\pi = 1.0$ GHz [20] the shear viscosity coefficient $\eta$ for alumina of 0.001 Pa·s was obtained.

Starting from the equation (3) it is not difficult to write all expressions which describe the interactions of viscoelastic spherical particles. For example, for linearized tangential force of "friction" we have

$$\frac{f_t}{E} = c_m a\delta + Ac_m a \frac{d\delta}{dt}, \qquad c_m = \frac{4}{(2-\nu)(1+\nu)},$$

where $\delta$ is the relative tangential displacement of contacting particles, $a = \sqrt{hd_g}/2$ is the contact spot radius, $h = d_g - r$ is the depth of particle overlapping.

The characteristic time $T = ((\pi\rho_m d_g^2)/(6E))^{1/2}$, which transforms the equations (1) to dimensionless form, is equal to 0.74 ps for our systems. The reduced time step of the numerical solving the equations (1) is $h_t = h_{t,d}/T = 0.04$. The relative decreases of model cell sizes $\Delta L_{\text{cell}}/L_{\text{cell}}$ corresponding to the time step are equal to values $10^{-5}$, $10^{-6}$, $10^{-7}$, $10^{-8}$, or $10^{-9}$. These values result in strain rates (in s$^{-1}$): $v_1 = 6.8\times10^8$, $v_2 = 6.8\times10^7$, $v_3 = 6.8\times10^6$, $v_4 = 6.8\times10^5$, and $v_5 = 6.8\times10^4$.

## 3. SIMULATION RESULTS AND DISCUSSION

Figure 1 presents the time-dependent hydrostatic pressure $p = \mathrm{Sp}(\sigma_{ij})/2$ averaged over 80 calculations and typical calculation curves for the compression rate $v_3$. Averaging the other rates has been performed over 100 ($v_1$, $v_2$) and 10 ($v_4$) independent calculations. For the rate $v_5$ the only one calculation has been carried out.

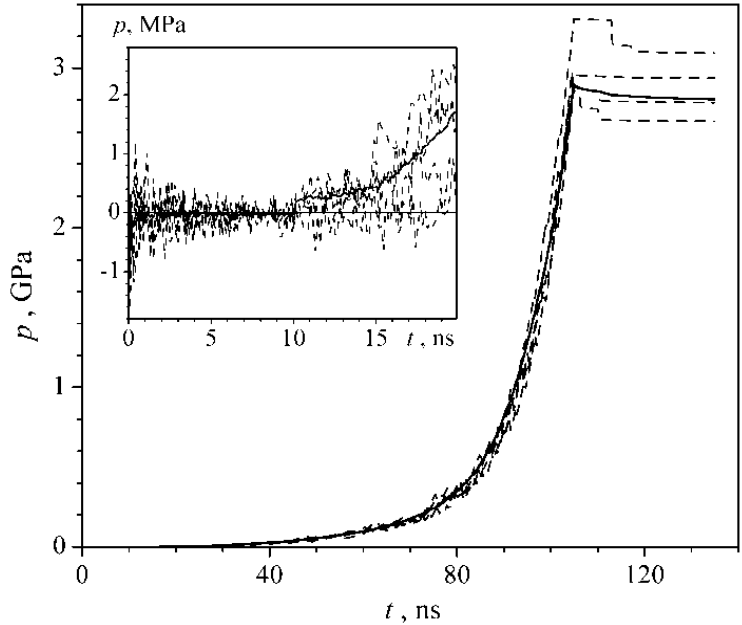

**Fig. 1.** The time dependence of pressure for the densification rate $v_3 = 6.8 \times 10^6$ s$^{-1}$. Dashed lines are examples of calculation curves, solid line is the average over 80 independent calculations. Insert shows the period of the preliminary relaxation (10 ns) and the beginning of the compression.

It is helpful to note that the initial structures are being generated by the algorithm [5], which places the neighboring particles at equilibrium distances when attraction compensates repulsion, requires however a preliminary relaxation step. It is needed since the dispersion forces between further particles are not taken into account in the algorithm that results in slight fluctuations of particles in the initial structure. In order to extinguish the fluctuations, initially generated structure relaxes for 10 ns (see the inset in Fig. 1).

Compression of the model cell was performed up to the density $\rho = 0.95$ where the pressure arrives at about 3 GPa. After that the system was relaxed during 30 ns. A considerable reduction of stresses is observed at this relaxation stage. This reduction for the hydrostatic pressure is well approx-

imated by an expression

$$p(t) = p_0 + p_1 \exp(-t / \tau_1) + p_2 \exp(-t / \tau_2) , \qquad (4)$$

Coefficients of the approximation (4) for the simulated strain rates are presented in Table 1. Post-compression relaxation proceeds in two stages: "rapid" with a characteristic time of about tenths of nanoseconds, and "slow", which lasts from several up to tens of nanoseconds. Change of hydrostatic pressure during relaxation decreases from 441 MPa (it is about 17% of the compaction pressure) at the densification rate $v_1$ almost to zero at the rate $v_5$. So, the compaction of the model system at the densification rate of the order of $10^4 \text{ s}^{-1}$ can be considered as a nearly quasistatic process.

**Table 1 –** The coefficients of approximations (4)

|  | $p_0$, MPa | $p_1$, MPa | $\tau_1$, ns | $p_2$, MPa | $\tau_2$, ns |
|---|---|---|---|---|---|
| $v_1$ | 2603.9 | 441.1 | 0.0719 | 35.3 | 4.9304 |
| $v_2$ | 2810.3 | 109.7 | 0.1977 | 72.0 | 4.3350 |
| $v_3$ | 2803.2 | 13.3 | 0.4486 | 84.1 | 9.2157 |
| $v_4$ | 2835.1 | 49.8 | 0.1515 | 29.0 | 10.0 |
| $v_5$ | 2868.6 | 0.0 | — | 1.31 | 30.0 |

Fig. 2 presents the compaction curves corresponding to the different densification rates. It is interesting that the $p(\rho)$ curve for $v_1$ rate has a local maximum at the beginning, which is very similar to the yield drop at stress-strain curves of metals [21]. This maximum has a dynamical nature and is caused by the retardation of relaxation processes from the powder compression. According to the simulation results, an increase in pressure up to the local maximum takes about 0.02 ns. This time is significantly less than the time of the "rapid" relaxation, which is about 0.07 ns (see Table 1) at the rate $v_1$. At slower densification rates this maximum disappears.

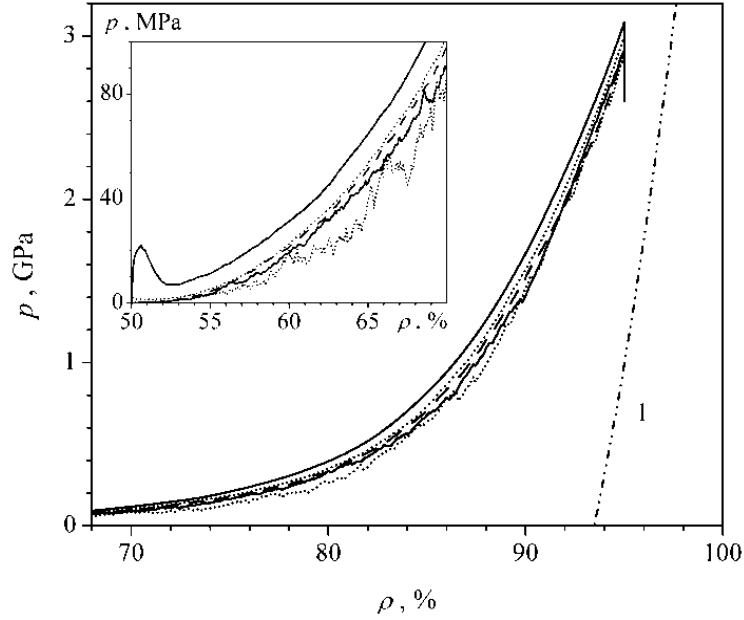

**Fig. 2.** Densification curves in "density – pressure" coordinates for strain rates $v_1$ (left solid line), $v_2$ (left dotted line), $v_3$ (dashed line), $v_4$ (right solid line), and $v_5$ (right dotted line). Line 1 is the asymptotic curve according to the eqs. (5) – (8). Insert shows the low pressure region in the expanded scale.

As one can see on the Fig. 2, all the densification curves can be adequately approximated in the limit of large densities and pressures. To obtain the asymptote we use the interrelation of regular packing density of the disks on the plane with mean coordination number $k_{av}$ (the number of particle contacts) in the form

$$\rho_{reg} = \frac{\pi / k_{av}}{\tan(\pi / k_{av})} \, . \tag{5}$$

At high pressures we can expect strong friction forces between particles. It should result to that the uniform compression of the system proceeds without relative displacement of particles. In such a case the density increases in 2D geometry as

$$\rho(h) = \frac{\rho_{reg}}{\left(1 - h / d_g\right)^2} \, . \tag{6}$$

For the hydrostatic pressure from Eq. (2) we have [9, 10]

$$p = \frac{\rho k_{av}}{\pi d_g^3} < f_n(h)(d_g - h) >, \qquad (7)$$

where $f_n$ is the normal part of the contact force without taking into account the viscous stresses, and angle brackets mean the average over the all pairs of interacting particles. Replacing $<\ldots>$ in Eq. (7) with the corresponding interaction laws we get the dependence of $p(\rho)$, which is implicitly determined by the equations $(5) - (7)$ where the mean coordination number $k_{av}$ is a parameter. According to the analysis of simulation results, the interrelation of coordination number $k_{av}$ with density for all the strain rates is well described by the expression

$$k_{av} = 2.3 + 3.0\rho^2. \qquad (8)$$

The asymptote relationship $p(\rho)$ determined by the equations $(5) - (8)$ is shown in Fig. 2 (curve 1). It can be seen that the densification curves $p(\rho)$ reach the asymptote in the high pressure limit. At the pressure of 3 GPa the error in the density according to the asymptote is less than 3%.

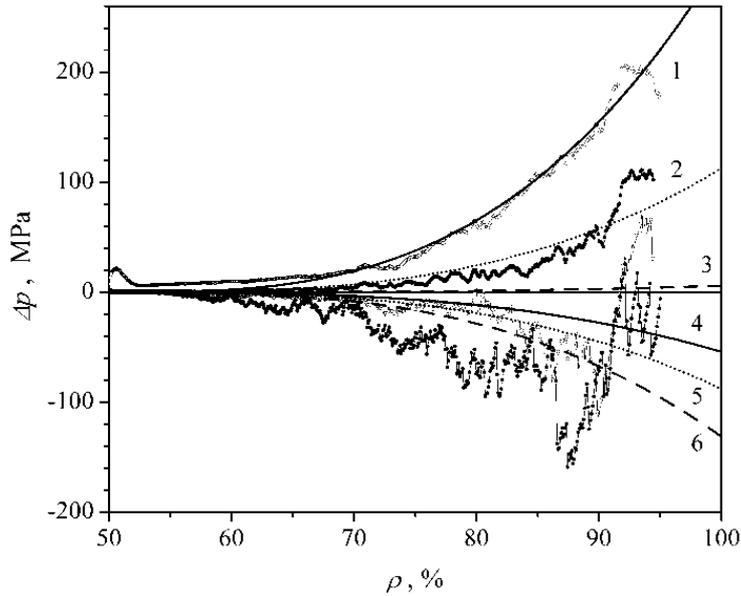

**Fig. 3.** The differences of compaction pressures from the pressure corresponding to the rate $v_3$ as a function of density. Symbols are the simulation results for rates $v_1$, $v_2$, $v_4$, and $v_5$ (from the top); smooth lines are the dependences of Eq. (9) for the densification rates $v_1 - v_5$ and for the quasistat-

ic conditions (the line 6).

To analyze the dependencies in Fig. 2, the densification curve $p(\rho)$ corresponding to the rate $v_3$ has been used as a reference one. Fig. 3 shows the differences between the other densification curves and the reference one. These differences are well approximated by the expression:

$$p(\rho, v) = p_3(\rho) + p_v(v)\left(\rho - \rho_0\right)^\gamma.$$

The analysis performed reveals that the index $\gamma \cong 3$, and the strain-rate-dependent coefficient $p_v$ is well described by the expression $p_v = p_{v0} + k_v v^{1/4}$. As a result, taking the quasistatic conditions ( $v \to 0$, the line 6 in Fig. 3) as a reference line we have obtained

$$p(\rho, v) = p_{\text{stat}}(\rho) + k_v v^{1/4}\left(\rho - \rho_0\right)^3. \qquad (9)$$

where $k_v = 21.5$ MPa s$^{1/4}$. Fig. 4 demonstrates the influence of compression rate on the acting pressure, which is determined by the equation (9).

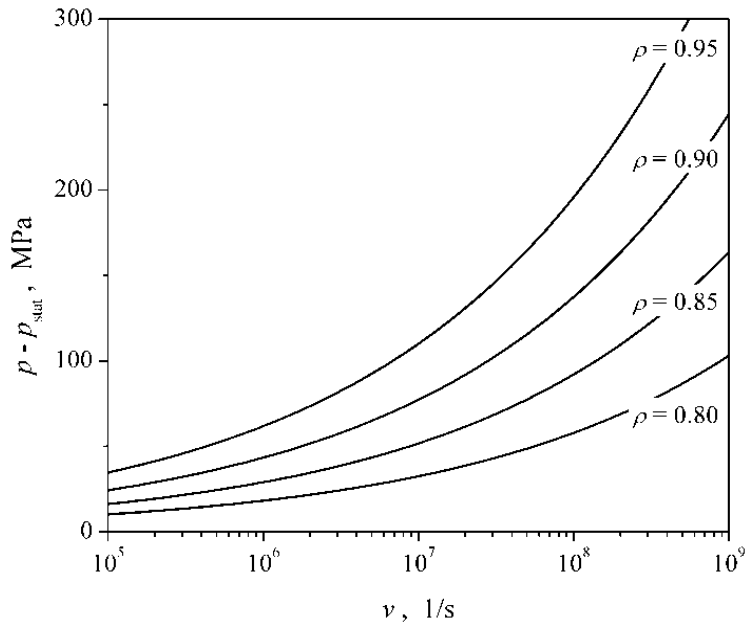

**Fig.** 4. Difference between dynamical and static compaction pressure at preset values of compact density as a function of strain rate.

For example, it can be seen from Fig. 4 that achievement of the density $\rho = 0.95$ with the strain rate $v = 10^8$ s$^{-1}$ requires the pressure, which is larger than that at quasistatic process by 200 MPa.

## 4. CONCLUSION

For the first time, the influence of the compression rate on the compactibility of oxide nanopowders has been studied by a two-dimentional granular dynamics method. Processes of stress relaxation after the stage of high-speed compression with the strain rates of $10^4 - 10^9$ s$^{-1}$ up to the relative density $\rho = 0.95$ have been analyzed. In particular, it has been found that the time of the stress relaxation is about 10 ns when the strain rate decreases down to $10^6$ s$^{-1}$. At this case the decrease of the mean pressure at the relaxation stage does not exceed 100 MPa which is significantly smaller than the compaction pressure (about 3 GPa). The explicit dependence of the compaction pressure, which is connected with the yield strength within phenomenology of powder body [1], on the strain rate has been established. It has been found that in 2D geometry, nanosized powders demonstrate the power-law dependence of pressure on strain rate as $p \propto v^{1/4}$, in contrast to plastic flow of metals, where the yield strength is proportional to the logarithm of the strain rate [12].

## ACNOWLEDGMENTS

The work has been fulfilled in the frame of the state task project No. 0389-2014-0002 and supported by RFBR (project No. 16-08-00277).